\documentclass[twocolumn,showpacs,preprintnumbers,amsmath,amssymb,floatfix]{revtex4}
\usepackage{epsf,epsfig,graphicx}

\newcounter{muni}

\begin{document}
\hbadness=10000 \pagenumbering{arabic}



\title{ Large Time-dependent CP Violation in $B_s^0$ System
        and Finite $D^0$-$\bar D^0$ Mass Difference
        in Four Generation Standard Model
}
\author{Wei-Shu Hou$^{a}$}
\author{Makiko Nagashima$^b$}
\author{Andrea Soddu$^{c}$}
\affiliation{ $^a$Department of Physics, National Taiwan
 University, Taipei, Taiwan 10617, R.O.C. \\
$^b$Physique des Particules, Universit\'e
 de Montr\'eal, 
 Montr\'eal, 
 Quebec, Canada H3C 3J7 \\
$^c$Department of Particle Physics, Weizmann Institute
 of Science, Rehovot 76100, Israel
}
\date{\today}

\begin{abstract}
Combining the measured $B_s$ mixing with $b \to s\ell^+\ell^-$
rate data, we find a sizable 4 generation $t'$ quark effect is
allowed, for example with $m_{t'} \sim 300$ GeV and
$V_{t's}^*V_{t'b} \sim 0.025\, e^{\pm i\, 70^\circ}$, which could
underly the new physics indications in CP violation studies of $b
\to s\bar qq$ transitions. With positive phase, large and negative
mixing-dependent CP violation in $B_s$ system is predicted, $\sin
2\Phi_{B_s} \sim -0.5$ to $-0.7$. This can also be probed via
width difference methods. As a corollary, the short distance
generated $D^0$-$\bar D^0$ mass difference is found to be
consistent with, if not slightly higher than, recent B factory
measurements, while CP violation is subdued with $\sin 2\Phi_D
\sim -0.2$.

\end{abstract}

\maketitle


Two decades after discovering large $B_d^0$-$\bar B_d^0$
mixing~\cite{dmdARGUS}, the long standing bound of 14.4
ps$^{-1}$~\cite{PDG} on $B_s^0$-$\bar B_s^0$ mixing finally turned
into a precise measurement~\cite{dmsCDF},
\begin{equation}
\Delta m_{B_s} = 17.77\pm 0.10 \pm 0.07\ {\rm ps}^{-1},
 \label{DmBs}
\end{equation}
and the focus is now on the associated $CP$ violation (CPV). The
measured~\cite{PDG} large CPV phase $\sin2\Phi_{B_d} \sim 0.7$ in
$B_d$ mixing (also called $\sin2\phi_1$ or $\sin2\beta$) is
consistent with the Standard Model (SM). However, $\sin
2\Phi_{B_s}^{\rm SM} \simeq -0.04$ is small, and offers a window
on New Physics (NP). The Tevatron can probe mixing-dependent CPV
in $B_s$ decay only if $\sin 2\Phi_{B_s}$ is large.
Another method using the width difference has recently
yielded~\cite{dgammaD0}
\begin{equation}
 \Delta\Gamma_{B_s} = 0.13 \pm 0.09\ {\rm ps}^{-1}, \
 \phi_s = -0.70^{+0.47}_{-0.39},
 \label{dgsphis}
\end{equation}
where $\phi_s \equiv 2\Phi_{B_s} \sim -40^\circ$ is sizable, but
only 1.2$\sigma$~\cite{dgammaD0} away from zero. But it is worth
emphasizing that {\it any evidence for $\sin 2\Phi_{B_s} \neq 0$
before the turning on of LHC would herald New Physics.}

The situation in $D^0$-$\bar D^0$ mixing is both similar and
different. There is now an indication for $\Delta m_D\neq
0$~\cite{Staric},
\begin{equation}
x_D \equiv \Delta m_{D}/\Gamma_D = 0.80\pm 0.29 \pm 0.17\ \%,
 \label{DmD}
\end{equation}
which is 2.4$\sigma$ from zero, while indications for effective
width differences have been further strengthened,
\begin{eqnarray}
y_{CP} &=& 1.31 \pm 0.32 \pm 0.25\ \%,
 \label{yCP} \\
y_D' &=& 0.97 \pm 0.44 \pm 0.31\ \%
 \label{yp}
\end{eqnarray}
at 3.2$\sigma$~\cite{yCPBelle} and 3.9$\sigma$~\cite{ypBaBar} from
zero, respectively.
With no evidence for CPV, $y_{CP} \cong y_D \equiv
\Delta\Gamma_D/2\Gamma_D$ is expected, while $y_D' = y_D\cos\delta
- x_D\sin\delta$ mixes $y_D$ and $x_D$ by a strong phase $\delta$
in $D\to K\pi$ decay. The SM predicts the {\it short distance}
$\Delta m_D^{\rm SD}$ and associated CPV $\sin2\Phi_D$ to be
negligible, but long distance effects could generate $y_D\sim$ \%,
which in principle could generate the observed $x_D$. But a finite
$x_D^{\rm SD}$ or $\sin2\Phi_D$ would indicate NP.

Can $\sin2\Phi_{B_s}$ be large? If so, can it be linked to finite
$\Delta m_{D}^{\rm SD}$? Can $\sin2\Phi_{D}$ be finite? Although
Eqs.~(\ref{DmBs})-(\ref{yp}) are not inconsistent with SM, the
window for NP is tantalizing. In this paper we show that, by
combining $\Delta m_{B_s}$ with $b\to s\ell^+\ell^-$ rate and
enlarging to 4 sequential generations (SM4), a sizable CPV phase
in $V_{t's}^*V_{t'b}$ is typically inferred, leading to sizable
$\sin 2\Phi_{B_s} < 0$. This new CPV phase is preferred by NP
hints in CPV in charmless $b\to s\bar qq$ transitions. SM-like CPV
in $b\to d$ transitions follow from imposing $Z\to b\bar b$ and
kaon constraints. Besides $\sin 2\Phi_{B_s} < 0$ and rather
enhanced $K_L\to \pi^0\nu\bar\nu$, we find $\Delta m_D^{\rm SD}$
at the level of Eq.~(\ref{DmD}) or higher, but $\sin2\Phi_D$ is
subdued.
The unifying thread behind all these phenomena is the
nondecoupling of $t$ and $t'$ (and $b'$) quarks, where $t'$ brings
in two new CPV phases.

The 4th generation is in fact disfavored by electroweak precision
tests, although loopholes do exist~\cite{PDG}. However, direct
search for $b'$ and $t'$ quarks has never ceased to be pursued at
the energy frontier~\cite{PDG}, which would open up greatly in the
era of LHC. One should therefore keep an open mind when searching
for NP in the flavor and CPV frontier. After all, with the
richness of phenomena already from the $3\times 3$ quark mixing
matrix $V$, enlarging it to $4\times 4$ implies considerable
enrichment. What we stress is that the 4th generation very
naturally impacts on box and electroweak penguin (EWP) diagrams
that enter $\Delta m_{B_s}$ and ${\cal B}(b\to s\ell^+\ell^-)$, by
the nondecoupling of $t$ and $t'$ quark effects~\cite{HWS}.
Destructive interference can allow a sizable $t'$ contribution in
association with a large CPV phase in
$V_{t's}^*V_{t'b}$~\cite{AHphase}.

$B_s$ mixing and $b\to s\ell^+\ell^-$ all involve $b
\leftrightarrow s$ transitions, where currently there are two
hints for NP involving CPV. Mixing dependent CPV measured in many
$b\to s\bar qq$ modes give~\cite{HFAG} the trend ${\cal S}_{s\bar
qq} < {\cal S}_{c\bar cs}$ (the $\Delta {\cal S}$ problem). A
second hint is the observed~\cite{HFAG} difference in direct CPV
in $B^0\to K^+\pi^-$ vs $B^+\to K^+\pi^0$ (the $\Delta {\cal
A}_{K\pi}$ problem). So the hope for large $\sin 2\Phi_{B_s}$ is
not unfounded. Fourth generation effects through EWP may also be
behind these NP hints, as has been shown in several
papers~\cite{Kpi0HNS,HNRS,HLMN}.
In fact, in Ref.~\cite{Kpi0HNS} we showed that the parameter space
implied by $\Delta {\cal A}_{K\pi}$ is independently supported by
combining the $\Delta m_{B_s}$ bound at that time with ${\cal
B}(b\to s\ell^+\ell^-)$. However, since $\Delta m_{B_s}$ and
${\cal B}(b\to s\ell^+\ell^-)$ suffer far less hadronic
uncertainties, with $\Delta m_{B_s}$ now precisely known, the
strategy should be switched around.

Throughout this paper, we will take $m_{t'} \simeq 300$ GeV for
sake of illustration. We stress that {\it changing $m_{t'}$ does
not affect the effects that we discuss, but leads to a shift in
the parameter range}.
Extending the quark mixing matrix $V$ from $3\times 3$ to $4\times
4$, all processes involving flavor are affected. With
$V_{t's}^*V_{t'b}$ determined from $b\to s$ transitions, we were
able to~\cite{globalHNS} more or less fix $V$ from considering
$Z\to b\bar b$ and rare kaon constraints. Remarkably, the
stringent kaon constraints imply~\cite{globalHNS} $b\to d$
measurements such as $\Delta m_{B_d}$ and $\sin2\Phi_{B_d}$ would
be SM-like, a necessary condition to survive NP
tests~\cite{fitter}, while $K_L \to \pi^0\nu\bar\nu$ could be
greatly enhanced.
It is critical to include (two) new CPV phases, not only because
they are naturally present, but also because it enriches and
enlarges~\cite{AHphase} the parameter space and phenomena. A
recent study~\cite{Louvain06} of the 4th generation, aimed towards
checking whether $V_{tb} = 1$, ignored CPV and assumed $V_{t's}$
to be real.

\begin{figure}[t!]
\smallskip  
\includegraphics[width=1.64in,height=1.1in,angle=0]{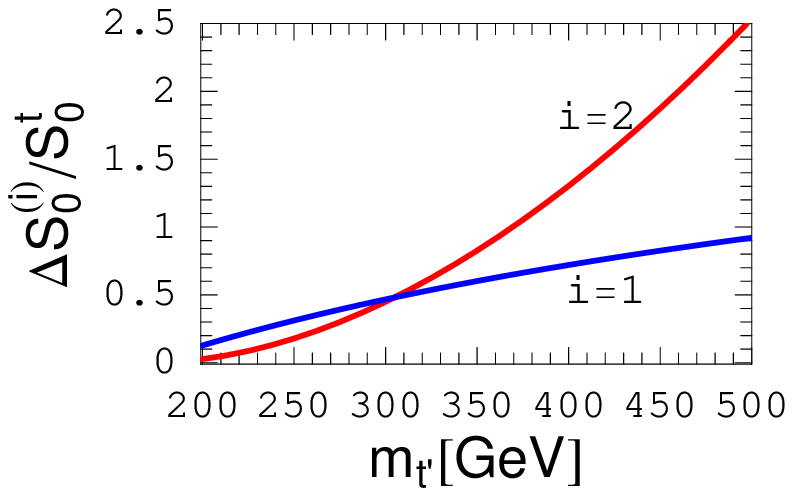}
\vspace{3mm}\hspace{-1mm}
\includegraphics[width=1.65in,height=1.11in,angle=0]{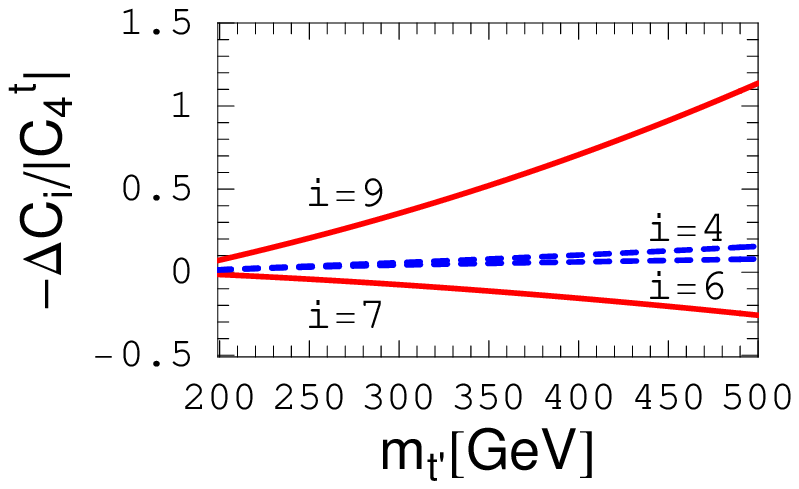}
\vskip-0.1cm
 \caption{
 Normalized $t^\prime$ corrections
 (a) $\Delta S_0^{(i)}/S_0^t$ (for box diagrams) and
 (b) $-\Delta\,C_i/|C_4^t|$ (for penguins)
  vs $m_{t^\prime}$.
 \label{fig:DeltaS0}
}
\end{figure}

Let us illustrate how the 4th generation enters box and EWP
diagrams.
For $b\to s$ transitions, 4 generation unitarity gives
$\lambda_u + \lambda_c + \lambda_t + \lambda_{t'} = 0$,
where $\lambda_i \equiv V_{is}^*V_{ib}$. Since $|\lambda_u| <
10^{-3}$ by direct measurement, one has
$
\lambda_t \cong -\lambda_c - \lambda_{t'}
$
$M_{12}$ for $B_s$ mixing is then proportional to
\begin{equation}
f_{B_s}^2 B_{B_s}
 \Bigl\{\lambda_c^2 S_0(t,t) - 2\lambda_c\lambda_{t'}
                               \Delta S_0^{(1)} 
 + \lambda_{t'}^2 \Delta S_0^{(2)}\Bigr\},
 \label{M12}
\end{equation}
where the first SM3 term is practically real ($\Gamma_{12}$, which
gives $\Delta\Gamma_{B_s}$, is generated by interference of $b\to
c\bar cs$ decay final states between $B_s$ and $\bar B_s$).
With
\begin{eqnarray}
\Delta S_0^{(1)} &\equiv& S_0(t,t') - S_0(t,t),
 \nonumber \\
\Delta S_0^{(2)} &\equiv& S_0(t',t') - 2S_0(t,t') + S_0(t,t),
 \nonumber
\end{eqnarray}
the $t'$ terms in Eq.~(\ref{M12}) respect GIM cancellation and
vanish with $\lambda_{t'}$, analogous to $\Delta\,C_i \equiv
C^{t^\prime}_{i} - C^t_{i}$~\cite{Kpi0HNS,AHphase} that modifies
the Wilson coefficients $C_i$ for $b\to s\bar qq$ (and
$s\ell^+\ell^-$) decays.
The normalized $\Delta S_0^{(i)}$ (to $S_0^t = S_0(t,t)$) are
plotted in Fig.~1(a) vs $m_{t'}$, which can be compared to
$\Delta\,C_{7,9}$ (normalized to $C_4^t$~\cite{Kpi0HNS}) plotted
in Fig.~1(b). The stong $m_{t'}$ dependence illustrates the
nondecoupling of SM-like heavy quarks from box and EWP
diagrams~\cite{HWS}. In contrast, the strong penguin corrections
$\Delta\,C_{4,6}$ shown in Fig.~1(b) have very mild $m_{t'}$
dependence~\cite{HouQCDP}. The electromagnetic (i.e. photonic)
penguin is likewise, which we will return to in our discussion
later. Thus, the 4th generation is of particular interest for
processes involving boxes and $Z$ penguins, the focus of our
study.

Besides the strong $m_{t'}$ dependence of $B_s$ mixing,
$\lambda_{t'}$ brings in a weak phase, which we parameterize as
\begin{eqnarray}
\lambda_{t'} \equiv V_{t's}^*V_{t'b} \equiv r_{sb} e^{i\phi_{sb}}.
 \label{lamtp}
\end{eqnarray}
This phase was not emphasized 20 years ago in the original work of
Ref.~\cite{HWS}, leading some authors to claim very narrow allowed
parameter range from data. But as was emphasized~\cite{AHphase}
later, the phase enriches the parameter space considerably,
allowing {\it large} effects from $t'$ even when {\it both}
$\,\Delta m_{B_s}$ and ${\cal B}(b\to s\ell^+\ell^-)$ appear
SM-like. For example, when $\lambda_{t'}$ is close to imaginary,
the $t$ and $t'$ contributions are largely real and imaginary,
respectively, hence add only in quadrature and do not interfere in
${\cal B}(b\to s\ell^+\ell^-)$. But this is just ripe for CPV.

\begin{figure}[t!]
\smallskip  
\hspace{-0.2mm}
\includegraphics[width=1.652in,height=1.15in,angle=0]{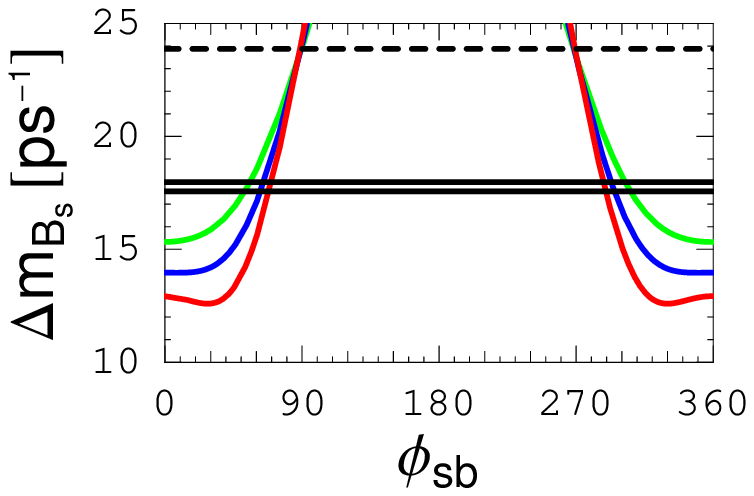}
\hspace{-3mm}
\includegraphics[width=1.68in,height=1.163in,angle=0]{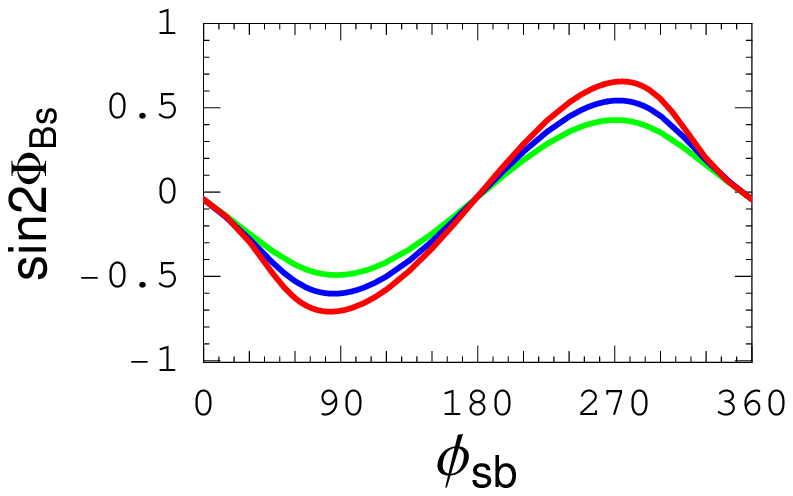}
\vspace{3mm}\hspace{0.5mm}
\includegraphics[width=1.67in,height=1.16in,angle=0]{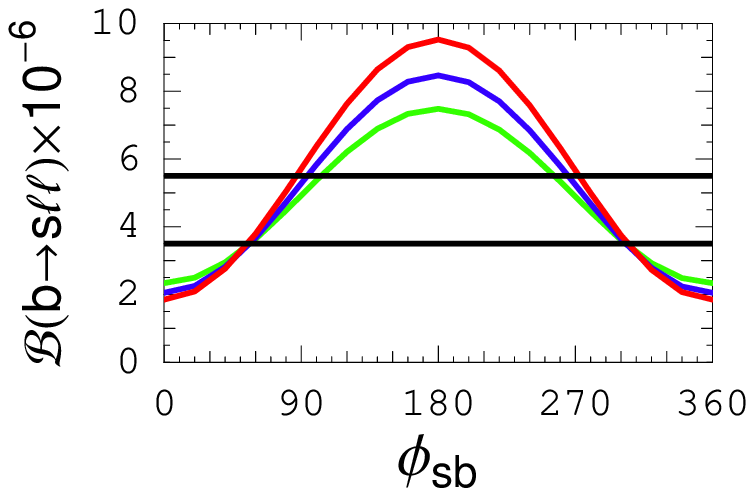}
\vspace{-0.5mm}\hspace{-3mm}
\includegraphics[width=1.655in,height=1.15in,angle=0]{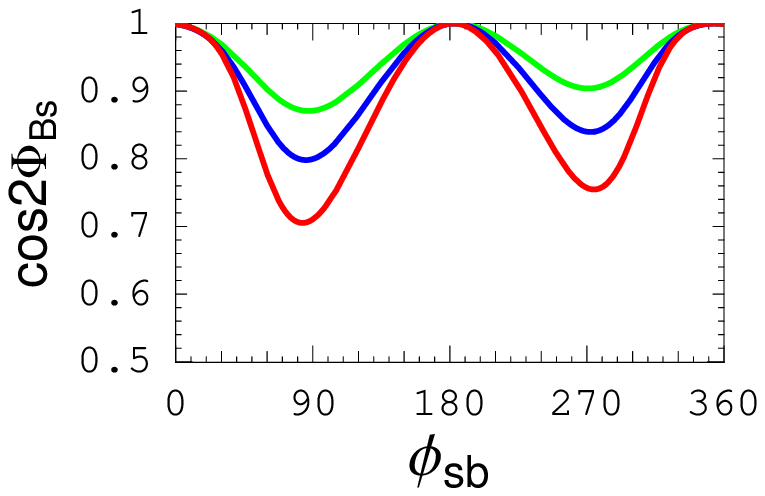}
\vspace{6.mm}
\caption{
 (a) $\Delta m_{B_s}$,
 (b) $\sin2\Phi_{B_s}$,
 (c) ${\cal B}(b\to s\ell^+\ell^-)$ and
 (d) $\cos2\Phi_{B_s}$
 vs $\phi_{sb}$, for $m_{t^\prime}=300$ GeV and
 $r_{sb}=$ 0.02, 0.025 and 0.03,
 where $V_{t^\prime s}^\ast V_{t^\prime b} \equiv r_{sb} \, e^{i\phi_{sb}}$.
 Larger $r_{sb}$ gives stronger variation.
 The dashed line in (a) is the nominal SM3 result, and the solid bands for
 (a) and (c) are $2\sigma$ and $1\sigma$ experimental ranges, respectively.
}
 \label{fig:btos}
\end{figure}

In the following, we take $m_t=170$ GeV. The central value of
$f_{B_s}\sqrt{B_{B_s}} = 295 \pm 32$ MeV \cite{latHPQCD} would
give the SM3 expectation of $\sim 24$ ps$^{-1}$~\cite{foot1},
which seems a little high compared with Eq.~(\ref{DmBs}). Of
course, $f_{B_s}\sqrt{B_{B_s}}$ could be in the lower range, but
{\it it could also turn out higher}. Some New Physics that
interferes destructively with SM3 is clearly welcome. Keeping
$f_{B_s}\sqrt{B_{B_s}} = 295$ MeV,
we plot $\Delta m_{B_s}$ vs $\phi_{sb}$ in Fig.~2(a) for $m_{t'}
=$ 300 GeV and several $r_{sb}$ values. The SM3 value is shown as
dashed line, and the 2$\,\sigma$ range
of Eq.~(\ref{DmBs}) is the solid band. We see that destructive
interference occurs for $\phi_{sb}$ in 1st and 4th quadrants,
bringing $\Delta m_{B_s}$ down from SM3 value to the CDF range,
while the 2nd and 3rd quadrants are ruled out. For given $r_{sb}$,
one projects a rather narrow $\phi_{sb}$ range. For $r_{sb} =$
0.02, 0.025, 0.03 (cf. $\lambda_c = V_{cs}^*V_{cb} \simeq 0.04$
for top contribution), we find $\phi_{sb} \simeq
52^\circ$--$55^\circ$, $62^\circ$--$64^\circ$,
$67^\circ$--$69^\circ$, respectively. These are quite imaginary
and implies large $\sin2\Phi_{B_s}$, which is plotted in
Fig.~2(b).

What is remarkable is the consistency of the observed $b\to
s\ell^+\ell^-$ rate with the above projection. This is because the
$b\to s\ell^+\ell^-$ process is also dominated by EWP and box
diagrams~\cite{HWS}. The current world average is ${\cal B}(b\to
s\ell^+\ell^-) = (4.5\pm 1.0)\times 10^{-6}$~\cite{PDG} (a cut on
$m_{\ell\ell} > 0.2$ GeV is applied), which is lower than two
years ago.
We follow the NNLO calculation of Ref.~\cite{Bobeth00}
and incorporate~\cite{Kpi0HNS} the 4th generation effect by
modifying short distance Wilson coefficients. Although formulas in
Ref.~\cite{Ali02}
are simpler, it is less clear how to incorporate the 4th
generation. We have checked that the $\Lambda_{\rm QCD}/m_b$
corrections in Ref.~\cite{Ali02} do not affect the NNLO result by
much.
We find 
${\cal B}(b\to s\ell^+\ell^-)|_{\rm SM} \simeq 4.26\times
10^{-6}$, in good agreement with data.

We plot ${\cal B}(b\to s\ell^+\ell^-)$ vs $\phi_{sb}$ in
Fig.~2(c), together with the $1\,\sigma$ range of data. The
dependence on $r_{sb}$ and $\phi_{sb}$ resembles the $\Delta
m_{B_s}$ case. We see that for $r_{sb} \sim$ 0.02, 0.025 and 0.03,
$|\phi_{sb}| \gtrsim 55^\circ$ is implied, which practically rules
out the allowed range from $\Delta m_{B_s}$ for $r_{sb} \sim$
0.02. Of course, there are uncertainties in both the measurements
and the theory, but our discussion makes clear that combining
${\cal B}(b\to s\ell^+\ell^-)$ with $\Delta m_{B_s}$, one can
constrain further the allowed parameter space.

Returning to Fig.~2(b), we see that $\sin2\Phi_{B_s} \sim -0.5$ to
$-0.7$ for $r_{sb} \sim 0.025$ to 0.03 for $\phi_{sb}$ in first
quadrant, which can give $\Delta {\cal A}_{K\pi} \sim 0.15$ and
$\Delta {\cal S} < 0$~\cite{HNRS,HLMN}. The latter excludes the
opposite sign case of $\phi_{sb}$ in fourth quadrant. Note that
one recovers the SM3 expectation of $\sim -0.04$ for $\phi_{sb} =
0$. Measuring mixing-dependent CPV should be no problem at LHC
experiments such as LHCb, but it is exciting that, with the
Tevatron sensitivity of 0.2 per experiment~\cite{reachTeV}, our
prediction could be tested at the 3$\sigma$ level before LHC turn
on.

We plot the somewhat redundant $\cos2\Phi_{B_s}$ in Fig.~2(d),
which can be probed by comparing $\Delta\Gamma_{B_s}/\Gamma_{B_s}$
measurements in non-CP and CP eigenstates~\cite{DFN01} and other
width difference approaches. The recent result by
D0~\cite{dgammaD0} is given in Eq.~(\ref{dgsphis}). The Belle
experiment has collected data in 2006 on the
$\Upsilon(5S)$~\cite{Drutskoy} for purpose of $B_s$ physics, but
it remains to be seen how much data would be needed for
$\cos2\Phi_{B_s}$ to be profitably probed. With central parameter
values we find $\sin2\Phi_{B_s} \sim -0.55$ hence $\cos2\Phi_{B_s}
\sim 0.84$. It is possible that the $\Delta\Gamma_{B_s}$ approach
could reach the precision to probe $\cos2\Phi_{B_s} \lesssim
0.85$.

\begin{figure}[t!]
\smallskip  
\includegraphics[width=1.65in,height=1.1in,angle=0]{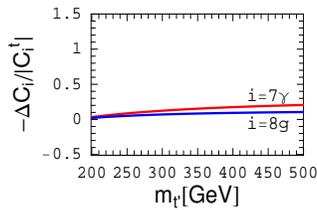}
\vskip-0.1cm
 \caption{
 $-\Delta C_{7\gamma}/|C_{7\gamma}^t|$ and
 $-\Delta C_{8g}/|C_{8g}^t|$ vs $m_{t^\prime}$,
 which governs on-shell photon and gluon emission.
 \label{fig:DeltaC7gamma}
}
\end{figure}

%
We have delayed the discussion of $b\to s\gamma$ until now, to
make clear the contrast: the well measured ${\cal B}(b\to
s\gamma)$, which is in agreement with SM expectation, does not
provide a better constraint. The relevant operator is
$\sigma_{\mu\nu}m_b R$, which is of dipole form, with coefficient
$C_{7\gamma}$, which should not be confused with $C_7$. The point
is that, unlike $C_9$ and $C_7$, the 4th generation correction
$\Delta C_{7\gamma}$ has rather mild dependence on $m_{t'}$. We
plot $-\Delta C_{7\gamma}$ normalized to $|C_{7\gamma}^t|$ (as
well as $-\Delta C_{8g}$ normalized to $|C_{8g}^t|$, the gluonic
dipole counterpart), in Fig.~3. Comparing with Fig.~1(b), the
$m_{t'}$ dependence is stronger than $-\Delta C_{4,6}/|C_4^t|$,
but much weaker than $-\Delta C_{7,9}/C_4^t$.

\begin{figure}[t!]
\smallskip  
\hspace{-0.2mm}
\includegraphics[width=3.63in,height=1.17in,angle=0]{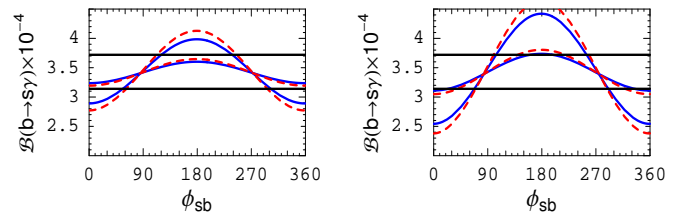}
%
\caption{
 ${\cal B}(b\to s\gamma)$ vs $\phi_{sb}$ for $m_{t^\prime}=$
  (a) 300 (solid), 350 (dashed), (b) 500 (solid), 700(dashed) GeV.
 For each $m_{t^\prime}$, two curves are shown, one each for
 $r_{sb}=$ 0.01 and 0.03, to illustrate the range, where larger
 value gives stronger variation.
}
 \label{fig:btosgamma}
\end{figure}

In addition to the milder dependence on $m_{t'}$, as we have
mentioned, $t$ and $t^\prime$ effects add mostly in quadrature for
the range of $\phi_{sb}$ under discussion. The combined effect is
that ${\cal B}(b\to s\gamma)$ does not pose a
problem~\cite{Kpi0HNS}. To make this point clear, we plot ${\cal
B}(b\to s\gamma)$ vs $\phi_{sb}$ in Figs.~4(a) and (b) for $m_{t'}
= 300$, 350, and 500, 700 GeV, respectively, for $r_{sb} = 0.01$
and 0.03, for sake of illustration. We have taken the
next-to-leading order results of Ref.~\cite{KN}, which give an SM
value of $3.4\times 10^{-4}$. Indeed the variation of the curves
are much milder compared with that of $b\to s\ell^+\ell^-$ plotted
in Fig.~2(c), even for extreme $m_t'$ masses such as 500--700 GeV.
The values stay closer to the experimental range, especially for
$\phi_{sb}$ near 90$^\circ$.
We also point out that the 4th generation in our scenario gives
rise to a smaller ${\cal A}_{\rm CP}(b\to s\gamma)$ than SM (see
Fig~3(c) of Ref.~\cite{Kpi0HNS}), and again does not pose any
problem.

We note that our $\phi_{sb}$ range tend to reduce ${\cal B}(b\to
s\gamma)$ by a small amount.Recent next-to-next-to-leading order
(NNL) calculations lead to a smaller~\cite{NNL} SM expectation.
This could pose a problem, but the NNL calculation is yet
incomplete~\cite{AEGG}. The situation should be watched, but in
any case this would be a generic problem, not just for the 4th
generation model.

A 4th generation affects all flavor changing phenomena, including
$D^0$ mixing. For this purpose, it is useful to take the $4\times
4$ parametrization~\cite{HSS87} that puts phases in $V_{ub}$ (as
in SM3), $V_{t's}$, and $V_{t'd}$.
With $V_{t's}^*V_{t'b}$ determined, in a previous
study~\cite{globalHNS} we used $Z\to b\bar b$ to fix $V_{t'b}\sim
-0.22$, hence $V_{t's} \sim -0.114\, e^{-i\,70^\circ}$. The kaon
constraints then imply $V_{t'd} \sim -0.0044\, e^{-i\,10^\circ}$,
leading to enhanced $K_L \to \pi^0\nu\nu$~\cite{globalHNS}. It is
nontrivial that $b\to d$ observables become SM-like, while $b\to
s$ has large CPV, which can be seen by comparing Fig.~1(b) to 1(a)
of Ref.~\cite{globalHNS}: one can barely tell apart the SM4 $b\to
d$ quadrangle from the SM3 triangle, while the SM4 $b\to s$
quadrangle is {\it large} and distinct from the squashed SM3
``triangle".

By unitarity one then finds $V_{cb'} \sim 0.116\, e^{i\,66^\circ}$
and $V_{ub'} \sim 0.028\, e^{i\,61^\circ}$, giving
\begin{equation}
V_{ub'}V_{cb'}^* \equiv r_{uc}\, e^{-i\,\phi_{uc}} \sim +0.0033\,
e^{-i\,5^\circ},
 \label{lambp}
\end{equation}
which affects $c\to u$ transitions via $b'$ loops. With
$|V_{ub}V_{cb}| \lesssim 10^{-4}$ by direct measurement hence
$
V_{ud}V_{cd}^* + V_{us}V_{cs}^* + V_{ub'}V_{cb'}^* \cong 0,
$
Eq.~(\ref{lambp}) implies $V_{ud}V_{cd}^* \simeq -0.218$ and
$V_{us}V_{cs}^* \simeq 0.215$ are real to better than 3 decimal
places, very much like in SM3. These govern $c\to u\bar dd$ and
$u\bar ss$ decays that generate $y_D$~\cite{Falk}, which in turn
could generate $|x_D^{\rm LD}| \sim |y_D|$ by dispersion
relations.

Long-distance effects are beyond our scope. Our interest is in the
$V_{ub'}V_{cb'}^*$ term. Though small, it cannot be ignored, since
$m_{b'} \sim m_{t'}$ by electroweak precision constraints hence is
very heavy. This can generate $x_D^{\rm SD}$, analogous to
Eq.~(\ref{M12}) for $\Delta m_{B_s}$, which would be vanishingly
small in SM3 because of $|V_{ub}V_{cb}|^2$ suppression.

%
\begin{figure}[t!]
\smallskip  
\includegraphics[width=3.6in,height=1.19in,angle=0]{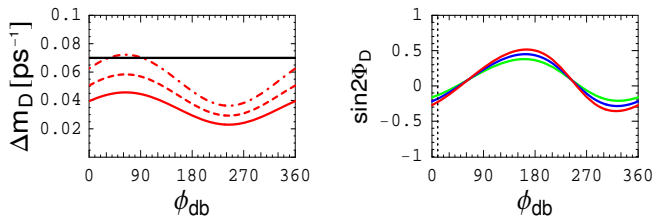}
%
\vskip-0.1cm \caption{
 (a) $\Delta m_D$ vs $\phi_{db}$
     for $m_{b'} =$ 230 (solid), 270 (dash) and 310 (dotdash) GeV
     and $r_{db} = 10^{-3}$, where
     $V_{t'd}V_{t'b}^* \equiv r_{db}\, e^{-i\,\phi_{db}}$;
 (b) $\sin2\Phi_D$ vs $\phi_{db}$ for $m_{b'} =$ 270 GeV and
     $r_{db} \sim (0.8,\; 1,\; 1.2)\times 10^{-3}$.
     The solid horizontal line in (a) is the PDG bound, while the
     dashed band is 2$\sigma$ range of Eq.~(\ref{DmD}).
 }
 \label{fig:Dmix}
\end{figure}

To illustrate the short-distance effect generated by $b'$, we take
$f_D\sqrt{B_{D}} = 200$ MeV and plot $\Delta m_D$ vs $\phi_{db}
\equiv \arg V_{t'd}^*V_{t'b}$ in Fig.~5(a), for $m_{b'} =$ 230,
270 and 310 GeV, respectively, and $r_{db} \equiv
|V_{t'd}^*V_{t'b}| = 10^{-3}$. The reason we plot against
$\phi_{db}$ is because $\phi_{db}$ and $r_{db}$ are
fitted~\cite{globalHNS} to kaon data, after fixing $r_{sb} \sim
0.025$ and $\phi_{sb} \sim 65^\circ$. We found $\phi_{db} \sim
10^\circ$ and $r_{db} \sim 10^{-3}$. We see that $\Delta m_D^{\rm
SD}$ lies just below the PDG bound (solid line), but is slightly
higher than the upper range of Eq.~(\ref{DmD}) (dashed band),
which is interesting. With $y_D \sim$ 1\%, $\vert x_D^{\rm
LD}\vert$ could be comparable~\cite{Falk}. Thus, our rough
prediction of $x_D^{\rm SD}$ could be plainly consistent with
Eq.~(\ref{DmD}), or imply destructive interference between short
and long distance $x_D^{\rm SD}$ and $x_D^{\rm LD}$. This is
reminiscent of the situation in $\Delta m_K$.

The critical test would be CPV. Although $\phi_{db} \sim
10^\circ$, in Fig.~5(b) we plot $\sin2\Phi_D$ vs $\phi_{db}$ for
$m_{b'} =$ 270 GeV and $r_{db} = (0.8,\; 1,\; 1.2)\times 10^{-3}$.
Because of the tiny $\phi_{uc} \sim 5^\circ$, our scenario {\it
predicts rather small, but still finite CPV} in $D^0$ mixing, at
not more than $-0.2$ level. It is remarkable that this is
consistent with no hints so far for CPV in $D$ mixing from
experiment~\cite{PDG,Staric,yCPBelle,ypBaBar}. But a nonzero
$\sin2\Phi_D$ should be of great interest in the longer term.

We note that experimental hints for $\Delta\Gamma_D$ has been
around for some time~\cite{PDG}, which stimulated the theoretical
suggestion~\cite{Falk} that $y_D\sim$ \% level is possible in SM.
The recent measurement of $y_D' = y_D\cos\delta - x_D\sin\delta$
by BaBar~\cite{ypBaBar} in wrong-sign $D^0\to K^+\pi^-$ decays,
Eq.~(\ref{yp}), is consistent with the previous Belle~\cite{Zhang}
measurement, though higher in value. Unfortunately, the strong
phase difference $\delta$ between the mixing and doubly-Cabibbo
suppressed amplitudes is unknown. With an active program at the B
factories and CLEO-c, and the start of BESIII and LHCb in 2008,
one expects the $y_D$ and $x_D$ measurements to improve.
Measurement of $\delta$, such as with methods developed in
Ref.~\cite{AsnerSun}, could shed further light on consistency of
different measurements. $y_{CP}$, $y_D$, $x_D$, $\delta$ could all
become measured in a few years. But with $y_D \sim 1\% \sim x_D$
the likely outcome, to elucidate whether one has SM or NP effect,
ultimately one has to measure $\sin2\Phi_D$, which perhaps could
be done by LHCb, but may have to await a Super B factory.


Comparatively, measuring $\sin2\Phi_{B_s} \sim -0.55$ before 2008
should be more promising. Note that the $\Delta\Gamma_{B_s}$
approach already yields some nonvanishing negative value,
Eq.~(\ref{dgsphis}), but the errors are still too accommodating.
With $\Delta m_{B_s}$ already measured, the standard approach
would be a time-dependent CPV study with more data, performing an
angular resolved simultaneously fit to both mass and width
mixings. Clearly the SM expectation of $-0.04$ is out of reach at
the Tevatron, but with 8~fb$^{-1}$ or more data and an expected
sensitivity of $-0.2$ per experiment~\cite{reachTeV}, it is of
great interest to see whether evidence for $\sin2\Phi_{B_s}\neq 0$
could emerge before LHC data arrives.


We have only illustrated the possible outcome for CPV in $B_s$
system, as well as mixing and CPV in $D^0$ system. The numbers are
not very precise and should not be taken literally, since
uncertainties in hadronic parameters such as $f_{B_s}^2B_{B_s}$
are large. But if a 4th generation is present in Nature, it is
quite likely that it contributes to $B_s$-$\bar B_s$ and
$D^0$-$\bar D^0$ mixings as well as $b\to s\ell^+\ell^-$ the way
we suggested. This is especially true, in light of the
mixing-dependent and direct CPV ``anomalies" in $b\to s\bar qq$
decays. Put in other words, the discovery of $\sin2\Phi_{B_s} < 0$
would provide more confirmation, as well as information on, the
scenario that we have proposed.
We stress once again that, changing $m_{t'}$ from our nominal
value of 300 GeV does not affect the effects that we discuss, but
leads to just a shift in the parameter range.

In summary, we show that the current measurements of $B_s$ mixing
and $b\to s\ell^+\ell^-$ rate, though consistent with the Standard
Model, can accommodate a four generation. The predicted large and
negative $CP$ violation phase in $B_s$ mixing can be tested
already at the Tevatron. As a corollary, we predict short distance
$D^0$ mixing to be close to present sensitivities, with subdued
but finite $CP$ violation phase that can be probed in the future.

\vskip 0.3cm \noindent{\bf Acknowledgement}.\
 This work is supported in part by NSC 94-2112-M-002-035 and
NSC 94-2811-M-002-053 of Taiwan, and HPRN-CT-2002-00292 of Israel.


\begin{thebibliography}{99}
\bibitem{foot1}
  Our number is higher than Ref.~\cite{fitter}, probably because
  these authors used lower $f_{B_s}\sqrt{B_{B_s}}$,
  and a slightly lower $m_t$ than ours.
\end{thebibliography}
\end{document}